\documentclass[11pt]{article}
\usepackage{a4wide}                       % to get a wider page margen
\usepackage[dvips]{graphicx}
\usepackage{epsf}                        % to make figures
\usepackage{latexsym}                    % to make fancy symbols
\usepackage{amsfonts}                    % to make fancy symbols
\usepackage{amssymb}                     % to make fancy symbols
\def\be{\begin{equation}}
\def\ee{\end{equation}}
\def\bea{\begin{eqnarray}}
\def\eea{\end{eqnarray}}

\def\part{\partial}

\def\R{\ensuremath{\mathbb{R}}}

\def\makeatletter{\catcode`\@=11}% 11:letter
\makeatletter
\def\mathbox#1{\hbox{$\m@th#1$}}%
\def\math@ccstyles#1#2#3#4#5#6#7{{\leavevmode
      \setbox0\mathbox{#6#7}%
      \setbox2\mathbox{#4#5}%
      \dimen@ #3%
      \baselineskip\z@\lineskiplimit#1\lineskip\z@
      \vbox{\ialign{##\crcr
             \hfil \kern #2\box2 \hfil\crcr
             \noalign{\kern\dimen@}%
             \hfil\box0\hfil\crcr}}}}
\def\mathaccstyles{\math@ccstyles\maxdimen}
\def\maththroughstyles{\math@ccstyles{-\maxdimen}}
\def\unity%
 {\maththroughstyles{.45\ht0}\z@\displaystyle {\mathchar"006C}\displaystyle 1}
\begin{document}

\rightline{FFUOV-08/12}
\vspace{1.5truecm}

%%%%%%%%%%%%%%%%%
\centerline{\LARGE \bf Confinement and Non-perturbative Tachyons}
\vspace{.5cm}
\centerline{\LARGE \bf in Brane-Antibrane Systems: A Summary\footnote{Contribution to the proceedings of the 4th EU RTN Workshop held in Varna (Bulgaria), 11-17 september 2008.}}
\vspace{1.3truecm}

\centerline{
    {\large \bf Norberto Guti\'errez}\footnote{E-mail address:
                                   {\tt norberto@string1.ciencias.uniovi.es}} 
    {\bf and}
   {\large \bf Yolanda Lozano}\footnote{E-mail address:
                                  {\tt ylozano@uniovi.es}}
    }
                                                            
\vspace{.4cm}

\centerline{{\it Departamento de F{\'\i}sica,  Universidad de Oviedo,}}
\centerline{{\it Avda.~Calvo Sotelo 18, 33007 Oviedo, Spain}}

\vspace{1truecm}

%%%%%%%%%%%%%%%%%
\centerline{\bf ABSTRACT}
\vspace{.5truecm}

\noindent
We present a worldvolume effective action suitable for the study of the
confined phase of a $(Dp,\bar{Dp})$ system at weak coupling, and we identify the mechanism by which the fundamental string arises from this action when the $Dp$ and the $\bar{Dp}$ annihilate. We  construct an explicit dual action, appropriate for the strong coupling regime, which realizes a generalized Higgs-St\"uckelberg phase for the (relative) (p-2)-form dual to the (overall) BI vector, the mechanism put forward by Yi and collaborators for realizing non-perturbatively the breaking of the overall $U(1)$ gauge group.
Our results  provide an explicit realization of the perturbative breaking of the overall $U(1)$ 
in a way that is consistent with the duality symmetries of String Theory.\\%{\it PACS:} 11.25.-w; 11.27.+d\\
%{\it Keywords:} Branes; Duality; Supergravity

\newpage

\section{Introduction}

$D\bar{D}$ systems have been widely used in the literature in the study of string theory in time dependent backgrounds, and more recently in the study of chiral symmetry breaking in holographic models of QCD.
A superposition of a D$p$-brane and an anti-D$p$-brane 
constitutes a non-BPS
system whose instability manifests itself in the existence of a complex
tachyonic mode in the open
strings stretched between the pair \cite{Sen4}. If when the tachyon rolls down to its true minimum its phase acquires a winding number,  because of its coupling to the relative U(1) vector field, a magnetic vortex soliton is created. This vortex solution carries D$(p-2)$-brane charge,
 as inferred
from the coupling $\int_{R^{p,1}}C_{p-1}\wedge dA^-$
 in the Chern-Simons action of the
($Dp$,$\bar{Dp}$), and charge conservation implies that a $D(p-2)$-brane
is left as a topological soliton.
In this process the relative $U(1)$
vector field acquires a mass through the Higgs mechanism by eating the phase of the
tachyonic field, and is removed from the low energy spectrum. The 
overall U(1) vector field, under which the tachyon is neutral, 
remains however unbroken, posing a puzzle \cite{Sred,Witten,Yi}.

It was suggested in \cite{Yi} that the overall $U(1)$ is in the confined phase, due to a dual Higgs mechanism in which magnetically charged 
tachyonic states 
associated to open $D(p-2)$-branes stretched
between the $Dp$ and the $\bar{Dp}$ condense. 
Evidence for such a situation comes from the
M-theory description of a $(D4,\bar{D4})$ system as an $(M5,\bar{M5})$ pair wrapped in the eleventh
direction. In this description an M2-brane stretched between the pair must contain a complex tachyonic excitation, and an M2-brane should emerge as well as the remaining topological soliton, after the $M5$ and the $\bar{M5}$ annihilate through a generalized Higgs mechanism. Since it is possible to perform the reduction down to Type IIA along a worldvolume direction of the $(M5,\bar{M5})$ which is either longitudinal or transverse to the stretched M2-brane, the $(D4,\bar{D4})$ pair that is obtained is either described, perturbatively, in terms of open string degrees of freedom, or, non-perturbatively, in terms of open $D2$-brane degrees of freedom. By T-duality the non-perturbative description for a $(Dp,\bar{Dp})$ system is in terms of open $D(p-2)$-brane degrees of freedom. In this description the fundamental string arises through a dual Higgs mechanism \cite{Rey} in which magnetically charged tachyonic states associated to the open $D(p-2)$-branes stretched between the pair condense. The localized magnetic flux at strong coupling is then translated at weak coupling into a confined overall U(1) electric flux. 
This mechanism is however intrinsically non-perturbative, and this makes this description highly heuristic.

The explicit action that describes the dual Higgs mechanism at strong coupling has not been constructed in the literature, although qualitative arguments pointing at some particular couplings have been given \cite{Yi,BHY,GHY}.
The possibility of describing the region of vanishing tachyonic potential 
in terms of the $(p-2)$-form fields dual to the BI
vector fields was addressed in \cite{GHY}, and although the explicit dual action was
not constructed, it was argued that the dual Higgs mechanism proposed in \cite{Yi} could 
be realized if this action was the one associated to an Abelian Higgs model for
the relative $(p-2)$-form dual field. The
fundamental string would then arise as a Nielsen-Olesen solution. One of the results that 
we will present in these proceedings will be the construction of the explicit dual action 
in terms of the dual potentials. We will show however that the dual of this action does not describe an Abelian Higgs model for the relative $(p-2)$-form potential, contrary to the expectation in \cite{GHY}. The desired model will instead arise from a generalization of Sen's action from which we will be able to describe the confining phase for the overall U(1) at weak coupling.

\section{The $(Dp,\bar{Dp})$ system in dual variables}

Although the complete effective action describing a brane-antibrane pair has not been derived from first principles, it is known to satisfy a set of consistency conditions \cite{Sen3}. In the context of our discussion in these proceedings this action describes the Higgs phase for the relative BI vector field. We will work to second order in
$\alpha^\prime$, ignoring tachyonic couplings to $C_{p-1}$ and taking the other RR potentials to zero. 
This truncated action will contain however the relevant couplings for describing the most important aspects of the dynamics of the $(Dp,\bar{Dp})$ system, both in the Higgs and in the confining phases\footnote{Once it is extended as we do in next section in order to incorporate the non-perturbative degrees of freedom associated to the $(p-3)$-brane topological defects.}. Our starting point is then the action: 
\begin{eqnarray}
\label{Higgsinitial}
&&S(\chi,A)=\int d^{p+1}x\, \Bigl\{ e^{-\phi}
\Bigl(\frac12 F^++B_2\Bigr)\wedge *
\Bigl(\frac12 F^+ +B_2\Bigr)+\frac14 e^{-\phi} F^-\wedge * F^- +\nonumber\\
&&+ |T|^2 (d\chi -A^-)\wedge *
(d\chi -A^-)
+ d|T|\wedge * d|T| -V(|T|) +C_{p-1}\wedge F^-\Bigr\}\, .
\end{eqnarray}
Here we have set $2\pi\alpha^\prime=1$,
$A^+$ and $A^-$ are the overall and relative BI vector fields: $A^+=A+A^\prime$, $A^-=A-A^\prime$, and the
complex tachyon is parametrized as $T=|T|e^{i\chi}$. $V(|T|)$ is the tachyon potential \cite{Sen2}, whose precise form will be irrelevant for our analysis. The pullbacks of the spacetime fields into the worldvolume are implicit. 

The last coupling shows that when the tachyon condenses in a vortex-like configuration a $D(p-2)$-brane is generated as a topological soliton \cite{Sen4}.
 In this process the relative $U(1)$ vector field eats the
scalar field $\chi$, gets a mass and is removed from the low energy spectrum.
The overall
$U(1)$ vector field, under which the tachyon is neutral, remains unbroken,  but it
is believed to be confined \cite{Yi,Seneffac,Sen2,BHY}.

\subsection{The duality construction}

Note that since $A^{-}$ is massive it cannot  be dualized in the standard way.  We can however 
 use the standard procedure to dualize the phase of the tachyon and $A^+$. These fields are dualized, respectively, into a $(p-1)$-form, $W_{p-1}$, and a $(p-2)$-form, that we denote by $A_{p-2}^-$ given that due to the opposite orientation of the antibrane the relative and overall gauge potentials should be interchanged under duality. The intermediate dual action that is obtained after these two dualizations are carried out is such that, up to a total derivative term, $A^-$ becomes massless and can therefore be dualized in the standard way into $A_{p-2}^+$.
 
The final dual action reads:
\begin{eqnarray}
\label{dualinitialHiggs}
&&\hspace{-0.2cm}S(W_{p-1}, A_{p-2})=\int d^{p+1}x\Bigl\{ e^\phi \Bigl( \frac12 F^{+}_{p-1}+W_{p-1}+C_{p-1}\Bigr)\wedge * \Bigl( \frac12 F^{+}_{p-1}+W_{p-1}+C_{p-1}\Bigr)\nonumber\\
&&\hspace{-0.3cm}+ \frac14 e^\phi F^{-}_{p-1}\wedge * F^{-}_{p-1}+\frac{1}{4|T|^2}dW_{p-1}\wedge * dW_{p-1}
+d|T|\wedge * d|T|-V(|T|)-B_2\wedge F^{-}_{p-1}\Bigr\}
\end{eqnarray}

The action (\ref{dualinitialHiggs}) is an extension of the actions proposed in \cite{QT}, and it will become clear later that it
describes the confining phase for the overall $(p-2)$-form dual
potential. 
This phase arises after the condensation of zero-dimensional
topological 
defects which originate from the end-points of open strings stretched
between the branes. The interpretation of the low energy mode  $W_{p-1}$
is as describing the fluctuations of these defects, and is such that away from
the defects $W_{p-1}=dA_{p-2}^+$.
It can be seen that the original gauge invariance has been mapped into a gauge transformation of 
$W_{p-1}$ and $A_{p-2}^+$. This symmetry can be
gauge fixed by absorbing 
$F_{p-1}^{+}$ into $W_{p-1}$, which becomes then massive. 
The overall $A_{p-2}^{+}$ gauge potential is then removed
from the low 
energy spectrum through the so-called Julia-Toulouse mechanism \cite{JT}, which we will discuss further in the next section and is, essentially, the contrary of the 
more familiar Higgs mechanism. The Julia-Toulouse mechanism  is therefore the responsible for the removal of the relative U(1) at strong coupling. 
However it clearly sheds no light on the removal of $A^{+}$.

When comparing the action (\ref{dualinitialHiggs}) to the actions describing the confining phases of antisymmetric field theories presented in \cite{QT} one sees that the modulus of the tachyon plays the role of the density of condensing topological defects, as can be expected since the instability in the confining phase is originated by the presence of the topological defects. In the confining models of Quevedo and Trugenberger a consistency requirement is that the antisymmetric field theory in the Coulomb phase is recovered for zero density of defects. This is indeed satisfied by our action (\ref{dualinitialHiggs}) for vanishing tachyon, since the $|T|\rightarrow 0$ limit forces the condition that $W_{p-1}$ must be exact and can therefore be absorbed through a redefinition of $A^+$, recovering the Coulomb phase in dual variables.

Finally, as we will briefly mention in the next section, it is possible to see that
the $D(p-2)$-brane arises as 
a confined electric flux brane after the Julia-Toulouse mechanism \cite{GL}.

\section{Confinement at weak string coupling}

Quevedo and Trugenberger \cite{QT} made explicit in the framework of antisymmetric field theories
an old idea in solid-state physics due 
to Julia and
Toulouse \cite{JT}.
These authors argued that for a compact tensor field of rank $(h-1)$ in $(p+1)$-dimensions a 
confined phase might arise
after the condensation of $(p-h-1)$-dimensional topological defects.
The fluctuations of the continuous distribution
of topological defects generate a new low-energy
mode in the theory which can be described by a new $h$-form, 
$W_h$, such that away from the defects $W_{h}=dA_{h-1}$, where
$A_{h-1}$ is the original tensor field. 
The effective action describing the confining phase of the
antisymmetric tensor field then depends 
on a gauge invariant combination of the antisymmetric tensor field, $A_{h-1}$,
and the extended $h$-form, $W_{h}$. This
combination is such that when the density of topological defects vanishes
the original action describing the antisymmetric tensor field theory in the Coulomb phase is recovered. 
As discussed in \cite{QT}, the finite condensate phase is a natural generalization of the confinement phase for a four dimensional vector gauge field to arbitrary $(h-1)$-forms in $d$ dimensions.

Given that the worldvolume theory of a $(Dp,\bar{Dp})$ system is a vector field theory, the 
results in \cite{QT} for $h=2$ can be applied to this case. In this case the Coulomb phase is the phase with zero tachyon, and it is 
therefore described by the Lagrangian:
\begin{equation}
\label{Coulombinitial}
L(A)= e^{-\phi}
\Bigl(\frac12 F^{+} +B_{2}\Bigr)\wedge *
\Bigl(\frac12 F^{+} + B_{2}\Bigr)+\frac14 e^{-\phi} F^{-}\wedge * F^{-}
+C_{p-1}\wedge F^{-}\, .
\end{equation}

Developping now on the ideas in \cite{QT} we have that the
topological defects whose condensation will give rise to the confining phase are $(p-3)$-branes,
which originate in this case from the end-points of $D(p-2)$-branes stretched between
the $Dp$ and the $\bar{Dp}$. The new mode associated to the fluctuations of the defects is 
described by a 2-form, $W_{2}$, which will couple in the action through a gauge
invariant combination with the overall $U(1)$ vector field\footnote{This is forced by consistency with S- and T-dualities.}.
The action should depend as well on the density of topological defects, such that
when this density vanishes the original action in the Coulomb phase, given by (\ref{Coulombinitial}),
is recovered. We will see that, contrary to the actions constructed in \cite{QT}, where
the density of topological defects entered as a parameter which was 
interpreted as a new scale in the theory, in the $(Dp,\bar{Dp})$ case it must be a dynamical quantity
 because
it is related through duality to the modulus of the tachyonic excitation of the open $D(p-2)$-branes
in the dual Higgs phase. We will denote this field by 
$|\tilde{T}|$ and, moreover, we will use the duality with the Higgs phase to include in the action its 
kinetic and potential terms.

The action that we propose for describing the confining phase of the $(Dp,\bar{Dp})$ system is then given by:
\begin{eqnarray}
\label{confinitial}
S(W_{2},A)&=&\int d^{p+1} x \Bigl\{e^{-\phi}
\Bigl(\frac12 F^{+} +W_{2}+ B_{2}\Bigr)\wedge *
\Bigl(\frac12 F^{+} +W_{2}+ B_{2}\Bigr)
+\frac14 e^{-\phi} F^{-}\wedge * F^{-}+\nonumber\\
&&+\frac{1}{4|\tilde{T}|^2}
dW_{2}\wedge *dW_{2}+ 
d|\tilde{T}|\wedge * d|\tilde{T}| -V(|\tilde{T}|)+C_{p-1}\wedge F^{-}\Bigr\}\, .
\end{eqnarray}
This action has been constructed under four requirements.  
One is gauge invariance, both under gauge transformations of the
BI vector fields and under 
$W_{2}\rightarrow W_{2}+d\Lambda_{1}$, which ensures that only the gauge invariant part
of $W_{2}$ describes a new physical degree of freedom. This transformation must be 
supplemented by $A^{+}\rightarrow A^{+}- 2\Lambda_{1}$, a symmetry that has to be gauge
fixed.
The second  is
relativistic invariance. The third requirement is that the original action 
describing the Coulomb phase must be recovered when $|\tilde{T}|\rightarrow 0$. Indeed, when 
$|\tilde{T}|\rightarrow 0$ we must have that
$dW_{2}=0$, so that $W_{2}=d\psi_{1}$ for some 1-form $\psi_{1}$. This form
can then be absorbed by $A^{+}$, and the original action (\ref{Coulombinitial}) is recovered.
These requirements were the ones imposed in \cite{QT}. On the other hand, consistency with
the duality symmetries of string theory will later on imply that $W_{2}$ must couple only to the overall $U(1)$ vector field.

Now, in (\ref{confinitial}) 
$F^{+}$ can be absorbed by $W_{2}$, fixing the gauge symmetry,
and the action can then be entirely formulated 
in terms of $W_{2}$ and the relative vector field:
\begin{eqnarray}
\label{confinitial2}
S(W_{2},A^{-})&=&\int d^{p+1}x \Bigl\{e^{-\phi}
\Bigl(W_{2}+ B_{2}\Bigr)\wedge *
\Bigl(W_{2}+ B_{2}\Bigr)
+\frac14 e^{-\phi} F^{-}\wedge * F^{-}+\nonumber\\
&&+\frac{1}{4|\tilde{T}|^2}
dW_{2}\wedge *dW_{2}+ 
d|\tilde{T}|\wedge * d|\tilde{T}| -V(|\tilde{T}|)+C_{p-1}\wedge F^{-}\Bigr\}\, .
\end{eqnarray}
In this process the original gauge field $A^{+}$ has
been eaten by the new gauge field $W_{2}$, and has therefore been removed from
the low energy spectrum. This solves the puzzle of the unbroken overall $U(1)$ at weak string coupling through the Julia-Toulouse
mechanism.
Let us now see how the fundamental string arises from this action.

Consider first the $p=3$ case, which can be directly compared to the results in \cite{Sugamoto}.  In this case
the action (\ref{confinitial2}) is a generalization of the action proposed in \cite{Sugamoto} to describe the confining phase of a four dimensional Abelian gauge theory. This action was constructed as the dual of the four dimensional Abelian Higgs model, and it
allows a quantized electric vortex solution similar to the Nielsen-Olesen string. We see below that in our case this solution is identified as a fundamental string.

The construction of the vortex solution in \cite{Sugamoto} considers a non-vanishing 2-form vorticity source\footnote{In the construction in \cite{Sugamoto} the vorticity source is created by the phase component of the Higgs scalar of the original Abelian Higgs model. In our case it is created by the phase component of the tachyon field associated to open D-strings connecting the $D3$ and the $\bar{D3}$. This will become clear after the analysis in the next section.} along the $x^3$ axis,
and looks for a static and axially symmetric solution representing 
a static circulation of flow around the $x^3$ axis, 
satisfying the quantization condition
\begin{equation}
\label{elecflux}
\int_{D_\infty}e^3 ds=2\pi n\, ,
\end{equation}
where $D_\infty$ is a large domain in the $(x^1,x^2)$ plane including the origin. This solution corresponds to the Nielsen-Olesen string in the original Higgs model. As expected, the magnetic flux quantization condition has been mapped under duality onto an electric flux quantization condition, given by (\ref{elecflux}). The reader is referred to \cite{Sugamoto} for a more detailed discussion. For arbitrary $p$ it is easy to find a similar, generalized, electric vortex solution with the same properties.

Let us now see that the confined electric flux string solution corresponds in the $(Dp,\bar{Dp})$ case to the fundamental string. In this case we have an additional coupling
\begin{equation}
\int B_2\wedge *W_2
\end{equation}
in the effective action (\ref{confinitial2}), which shows that the quantized electric flux generates $B_2$-charge in the system.
Charge conservation then implies that the remaining topological soliton is the fundamental string.

As mentioned in the previous section, the $D(p-2)$-brane arises from the strongly coupled confining action (\ref{dualinitialHiggs}) derived in that section in a very similar way \cite{GL}. Therefore, the $D(p-2)$-brane arises either as a magnetic vortex solution after the Higgs mechanism at weak coupling or as confined electric flux brane after the Julia-Toulouse mechanism at strong coupling.

\section{Confinement at strong string coupling: The dual Higgs mechanism}

Inspired by Mandelstam-'t Hooft duality \cite{MT} we expect that the dual of the action (\ref{confinitial}) describes the Higgs phase for the $(p-2)$-form field dual to the overall BI vector. The dualization of the BI vector fields in (\ref{confinitial}) takes place in the standard way, given that they only couple through their derivatives. In turn, the 2-form $W_2$ is massive, but it can still be dualized in the standard way from the intermediate dual action that is obtained after dualizing the BI vector fields, in which it only couples through its derivatives. Let us call the dual of this form, a $(p-3)$-form, 
$\chi_{p-3}$. The final dual action reads:
\begin{eqnarray}
\label{Higgsdual}
&&S(\chi_{p-3},A_{p-2})=\int d^{p+1}x \Bigl\{e^\phi \Bigl(\frac12 F^{+}_{p-1}+C_{p-1}\Bigr)\wedge *
\Bigl(\frac12 F_{p-1}^++C_{p-1}\Bigr)
+\frac{1}{4} e^\phi F_{p-1}^{-}\wedge *
F_{p-1}^{-}\nonumber\\
&&+|{\tilde T}|^2 \Bigl( d\chi_{p-3}-A^{-}_{p-2}\Bigr)\wedge *
 \Bigl( d\chi_{p-3}-A^{-}_{p-2}\Bigr)+
d|{\tilde T}|\wedge * d|{\tilde T}| -V(|{\tilde T}|)-
B_{2}\wedge F_{p-1}^{-}\Bigr\}
\end{eqnarray}
where once again the overall and the relative gauge fields are interchanged. 

The action (\ref{Higgsdual}) describes an Abelian Higgs model for the relative $(p-2)$-form field, with the dual $(p-3)$-form $\chi_{p-3}$ playing the role of the associated Goldstone boson. That an effective mass gauge invariant term of this kind could drive the dual Higgs mechanism was suggested in \cite{Yi,BHY,GHY} (see also \cite{Rey}), although it could not be explicitly derived from the action describing the Higgs phase at weak coupling, i.e. from Sen's action. The Goldstone boson $\chi_{p-3}$ is associated to the fluctuations of the 
 $(p-3)$-dimensional topological defects that originate from the end-points of the $D(p-2)$-branes stretched between the $Dp$ and the $\bar{Dp}$. This is consistent with the fact that this field 
 is the worldvolume dual of the field $W_{2}$, which was accounting for these fluctuations in
 the confining action (\ref{confinitial}).
 Moreover,  we can identify for $p=3$ the condensing Higgs scalar as the modulus of the tachyonic mode associated to open D-strings stretched between the $D3$ and the $\bar{D3}$. Indeed when $p=3$  the action (\ref{Higgsdual}) turns out to be the S-dual of the original action (\ref{Higgsinitial}) describing the perturbative Higgs phase of the $(D3,\bar{D3})$ system. This is an important consistency check, 
although strictly speaking S-duality invariance would only be expected for zero tachyon.
In this duality relation
the modulus of the perturbative tachyon is mapped into $|{\tilde T}|$, which can then be interpreted as the modulus of the tachyonic excitation associated to the open D-strings. Since ${\tilde \chi}$ has also an interpretation as the phase of the dual tachyon we can think of 
${\tilde T}$ as the complex tachyonic mode associated to the D-strings stretched between the $D3$ and the $\bar{D3}$.  For $p\neq 3$, since the tachyonic condensing charged object is  a $(p-3)$-brane, 
the phase of the tachyon is replaced by a $(p-3)$-form. It would be interesting to clarify the precise way in which these fields arise as open $D(p-2)$-brane modes. 

Finally, if the brane and the antibrane annihilate through a generalized Higgs-St\"uckelberg mechanism in which $A_{p-2}^{-}$ gets a mass by eating the Goldstone boson $\chi_{p-3}$, we have that, if the Goldstone boson acquires a non-trivial winding number, $B_{2}$-charge is induced in the configuration 
through the coupling $\int_{\R^{p,1}}B_{2}\wedge F_{p-1}^{-}$ in (\ref{Higgsdual}). Charge conservation therefore implies that after the annihilation a fundamental string is left as a topological soliton. Since in this process the relative $(p-2)$-form field is removed from the low energy spectrum, and this field is dual to the original overall $U(1)$, this solves the puzzle of the unbroken $U(1)$, though through the mechanism suggested in \cite{Yi} which is intrinsically non-perturbative.

\section{Discussion}

As we have seen, a $(Dp,\bar{Dp})$ system admits two types of topological defects: particles and $(p-3)$-branes, which are, respectively, perturbative and non-perturbative in origin. The combined electric and magnetic Higgs mechanisms introduce mass gaps to both U(1) vector potentials, being the only remnants $D(p-2)$-branes and fundamental strings, realized as solitons on the common $(p+1)$-dimensional worldvolume. We have seen that it is possible to incorporate the non-perturbative degrees of freedom associated to the extended topological defects in the weak coupling regime, using Julia and Toulouse's idea, introducing a new form describing the fluctuations of these defects and imposing a set of consistency conditions. In fact, one can combine the weakly coupled action presented in section 3 with Sen's action in order to incorporate the degrees of freedom associated to both the zero dimensional and extended topological defects:
\begin{eqnarray}
\label{completa}
S(\chi,W_{2},A)&=&\int d^{p+1} x \Bigl\{e^{-\phi}
\Bigl(\frac12 F^{+} +W_{2}+ B_{2}\Bigr)\wedge *
\Bigl(\frac12 F^{+} +W_{2}+ B_{2}\Bigr)
+\frac14 e^{-\phi} F^{-}\wedge * F^{-}+\nonumber\\
&&+|T|^2(d\chi-A^-)\wedge * (d\chi-A^-)+d|T|\wedge * d|T|
+\frac{1}{4|\tilde{T}|^2}
dW_{2}\wedge *dW_{2}+\nonumber\\ 
&&+d|\tilde{T}|\wedge * d|\tilde{T}| -V(|T|)-V(|\tilde{T}|)+
C_{p-1}\wedge F^{-}\Bigr\}\, .
\end{eqnarray}
This action describes both 
 the perturbative and the non-perturbative Higgs mechanisms simultaneously at weak coupling, and
it admits both a magnetic vortex solution, which by charge conservation is identified with the $D(p-2)$-brane, and an electric vortex solution, identified as the fundamental string. 

\subsection*{Acknowledgements}

It is a pleasure to thank the organizers of the Varna meeting FU4. We would also like to thank Fernando Quevedo for useful discussions.
The work of N.G. was supported by a FPU Fellowship from the Spanish Ministry of Education. This work has been partially supported by the CICYT grant MEC-06/FPA2006-09199 and by the European Commission FP6 program MRTN-CT-2004-005104, in which the authors are associated to Universidad Aut\'onoma de Madrid.

\end{document}